\documentclass[longtitle,preprint,5p,times,twocolumn]{elsarticle}

\usepackage{amssymb}
\usepackage{amsmath}

\usepackage{caption}
\usepackage{xspace}
\usepackage{longtable}
\usepackage{pdflscape}
\usepackage{multirow}
\usepackage{ragged2e}
\usepackage{booktabs}
\usepackage{hyperref}

\usepackage[table]{xcolor}
\usepackage{comment}

\usepackage{lineno}

\journal{Nuclear Instruments and Methods A}

\begin{document}

\begin{frontmatter}


\title{Open Heavy Flavor Studies for the ECCE Detector at the Electron Ion Collider}

\def\theaffn{\arabic{affn}} 
%
%
%
%

\author[LANL]{X.~Li}
\author[MoreheadState]{J.~K.~Adkins}
\author[RIKEN,RBRC]{Y.~Akiba}
\author[UKansas]{A.~Albataineh}
\author[ODU]{M.~Amaryan}
\author[Oslo]{I.~C.~Arsene}
\author[MSU]{C. Ayerbe Gayoso}
\author[Sungkyunkwan]{J.~Bae}
\author[UVA]{X.~Bai}
\author[BNL,JLab]{M.D.~Baker}
% ...

%

% ...
\affiliation[BNL]{organization={Brookhaven National Laboratory},
	 city={Upton},
	 state={NY},
	 country={USA}} 
 
% ...
\affiliation[JLab]{organization={Thomas Jefferson National Accelerator Facility},
	 city={Newport News},
	 state={VA},
	 country={USA}} 
 
% ...
\affiliation[LANL]{organization={Los Alamos National Laboratory},
	 city={Los Alamos},
	 state={NM},
	 country={USA}} 
 
% ...
\affiliation[MoreheadState]{organization={Morehead State University},
	 city={Morehead},
	 state={KY},
	 }
 
% ...
\affiliation[MSU]{organization={Mississippi State University},
	 city={Mississippi State},
	 state={MS},
	 country={USA}} 
 
% ...
\affiliation[ODU]{organization={Old Dominion University},
	 city={Norfolk},
	 state={VA},
	 country={USA}} 
 
% ...
\affiliation[RIKEN]{organization={RIKEN Nishina Center},
	 city={Wako},
	 state={Saitama},
	 country={Japan}} 
 
% ...
\affiliation[RBRC]{organization={RIKEN BNL Research Center},
	 city={Upton},
	 state={NY},
	 country={USA}} 
	 
% ...
\affiliation[Sungkyunkwan]{organization={Sungkyunkwan University},
	 city={Suwon},
	 country={Republic of Korea}} 
 
% ...
\affiliation[UKansas]{organization={Unviersity of Kansas},
	 city={Lawrence},
	 state={KS},
	 country={USA}} 
 
% ...
\affiliation[Oslo]{organization={University of Oslo},
	 city={Oslo},
	 country={Norway}} 
 
% ...
\affiliation[UVA]{organization={University of Virginia},
	 city={Charlottesville},
	 state={VA},
	 country={USA}} 
 
% ...

\begin{abstract}
The ECCE detector has been recommended as the selected reference detector for the future Electron-Ion Collider (EIC). A series of simulation studies have been carried out to validate the physics feasibility of the ECCE detector. In this paper, detailed studies of heavy flavor hadron and jet reconstruction and physics projections with the ECCE detector performance and different magnet options will be presented. The ECCE detector has enabled precise EIC heavy flavor hadron and jet measurements with a broad kinematic coverage. These proposed heavy flavor measurements will help systematically study the hadronization process in vacuum and nuclear medium especially in the underexplored kinematic region.
\end{abstract}

\begin{keyword}
ECCE \sep Electron Ion Collider \sep Tracking \sep Heavy Flavor
\end{keyword}

\end{frontmatter}


\setcounter{tocdepth}{1}
\tableofcontents 

\section{Introduction}
\label{sec:intro}
The U.S. recently started the development of a new Electron-Ion Collider (EIC). Among others, the EIC science program \cite{Accardi:2012qut, NAP25171, AbdulKhalek:2021gbh} will study how colored quarks and gluons form colorless matter when propagating through space, a process known as hadronization. The future EIC will operate high luminosity and high energy $e+p$ and $e+A$ collisions with the instantaneous luminosity at $10^{33-34} cm^{-2}s^{-1}$ and a variety of center-of-mass energy from 20 GeV to 141 GeV. The nuclear species vary from proton to lead for different $e+p/A$ collision at the EIC. All these features will create an ideal environment to explore the hadronization process as a function of nuclear medium size.

The exploration of how colored quarks and gluons form into final state colorless hadrons is one of the primary physics drivers of the EIC. According to the factorization theorem \cite{Collins:1985ue}, inclusive hadron cross sections in $e+p$ and $e+A$ collisions contain both the initial state effects such as the (nuclear) parton distribution functions and final state effects such as the hadron fragmentation functions as defined in Eq. \ref{eq_Xsec},

\begin{equation}
\label{eq_Xsec}
  \sigma_{e+p/A} = f_{a}^{p/A}(x, Q^{2}) \otimes H_{a\gamma* \rightarrow c} \otimes D_{c}^{e+p/A}(z,\mu),
\end{equation}
where $f_{a}^{p/A}(x, Q^{2})$ is the parton distribution function for a parton with flavor $a$ and carrying the longitudinal momentum fraction $x$ and energy scale $Q^{2}$ inside a proton or a nucleus, $H_{a\gamma* \rightarrow c}$ is the parton hard scattering process which can be calculated by perturbative QCD, $D_{c}^{e+p/A}(z,\mu)$ is the fragmentation function for a produced parton with flavor $c$ that forms into a hadron with the energy/momentum fraction $z$ relative to this parton at the fragmentation scale $\mu$ in $e+p$ or $e+A$ collisions. The fragmentation function also depends on the parton flavor and is expected to vary in nuclear medium from vacuum. Due to their larges masses, heavy flavor quark fragmentation functions dominate in the kinematic region of $z > 0.4$, in which the fragmentation function is not fully constrained by existing experimental data. In the leading order QCD approach, heavy quark production at the EIC is mainly from the photon-gluon fusion process. Therefore, heavy quark production at the EIC can provide unique constraints on the accessed gluon parton distribution functions.

The advanced silicon vertex/tracking detector designed by the ECCE consortium \cite{ecce-paper-det-2022-03} enables a series of inclusive and differential heavy flavor hadron and jet measurements in $e+p$ and $e+A$ collisions. This paper focuses on extracting the nuclear medium effects on heavy quark fragmentation functions by comparing the reconstructed heavy flavor hadron cross sections between $e+p$ and $e+A$ collisions, through the measurements of heavy flavor hadron nuclear modification factor $R_{eA}$ defined below:

\begin{equation}
\label{eq3}
  R_{eA} = \frac{1}{A}\frac{\sigma_{e+A}}{\sigma_{e+p}},
\end{equation}
where $A$ is the nucleus mass number, $\sigma_{e+A}$ is the hadron/jet cross section in $e+A$ collisions and $\sigma_{e+p}$ is the hadron/jet cross section in $e+p$ collisions at the same collision energy. The projected nuclear modification factor $R_{eA}$ of reconstructed heavy flavor hadrons is studied as a function of hadron momentum fraction $z$, which is proportional to the ratio of heavy quark fragmentation function in $e+A$ collisions over that in $e+p$ collisions. The fragmentation functions in $e+p$ and $e+A$ collisions are integrated over the fragmentation scale $\mu$.

This paper summarizes the latest simulation studies for reconstructed heavy flavor hadrons and jets in $e+p$ collisions and the corresponding nuclear modification factor projections in $e+p$ and $e+Au$ collisions. The selected collision combination is 10 GeV electron and 100 GeV proton/nucleus, which is projected to reach the highest $e+A$ energy combination at the future EIC. The latest ECCE conceptual design has been implemented in the GEANT4 simulation. Heavy flavor reconstruction has been studied in PYTHIA~8 simulation with the latest EIC accelerator design and ECCE detector performance. The content includes the simulation configuration, heavy flavor reconstruction analysis procedure, physics projection and summary.
\section {Simulation Setup}
\label{sec:sim}
\subsection{Analysis Framework}
\label{sec:ana} The simulation framework for the heavy flavor hadron and jet reconstruction includes event generation in PYTHIA~8.2 \cite{Sjostrand:2014zea}, ECCE tracking detector performance extrapolated in GEANT~4 \cite{AGOSTINELLI2003250} through the Fun4All framework \cite{fun4allGithub, ecce-note-comp-2021-01}, parameterized detector performance of PID, EMCal and HCal discussed in section 3 and 11 in the EIC yellow report, estimated beam gas background from the hadron beam. As other beam backgrounds, such as the synchrotron radiation, are still under evaluation, we did not include these backgrounds in the simulation. In total 60 M PYTHIA events have been generated with the minimum $Q^{2}$ at $10~$Gev/$c^{2}$ and all charm and bottom hadron decay channels enabled. These high $Q^{2}$ events have been scaled with the associated minimum bias events for heavy flavor reconstructions. The PYTHIA 8 configuration setup is listed in Table~\ref{tab:py_set}.
 

\begin{table}[!htb]
\begin{center} 
\begin{tabular}{|c|c| } 
 \hline
Configuration parameter & Value \\ \hline 
PhaseSpace:Q2Min & 10 \\ \hline 
WeakBosonExchange:ff2ff(t:gmZ) & on \\ \hline
WeakBosonExchange:ff2ff(t:W) & on \\ \hline
SpaceShower:dipoleRecoil & on \\ \hline
heavy flavor hadorn decay (e.g., 413:onMode) & on \\ \hline 
 \end{tabular}
 \caption{The PYTHIA 8 simulation configuration for the heavy flavor hadron and jet reconstruction.}
    \label{tab:py_set}
\end{center}
\end{table}

Heavy flavor hadrons usually have shorter lifetime compared to light flavor hadrons. This results in a relatively short decay length ($100~\mu m < c\tau_{0} < 550~\mu m$) between the production of a heavy flavor hadron and its decay vertex in the rest framework. Reconstruction of heavy flavor hadrons and tagging a heavy flavor jet mainly rely on measuring particle displaced/secondary vertex or track Distance of Closet Approach (DCA), which is proportional to the particle decay length to search for the heavy flavor decay particles. Such measurements can be performed by a low material budget and high spatial resolution silicon vertex/tracking detector.

\begin{figure}[!ht]
\centerline{
\includegraphics[width=0.45\textwidth, angle = 0]{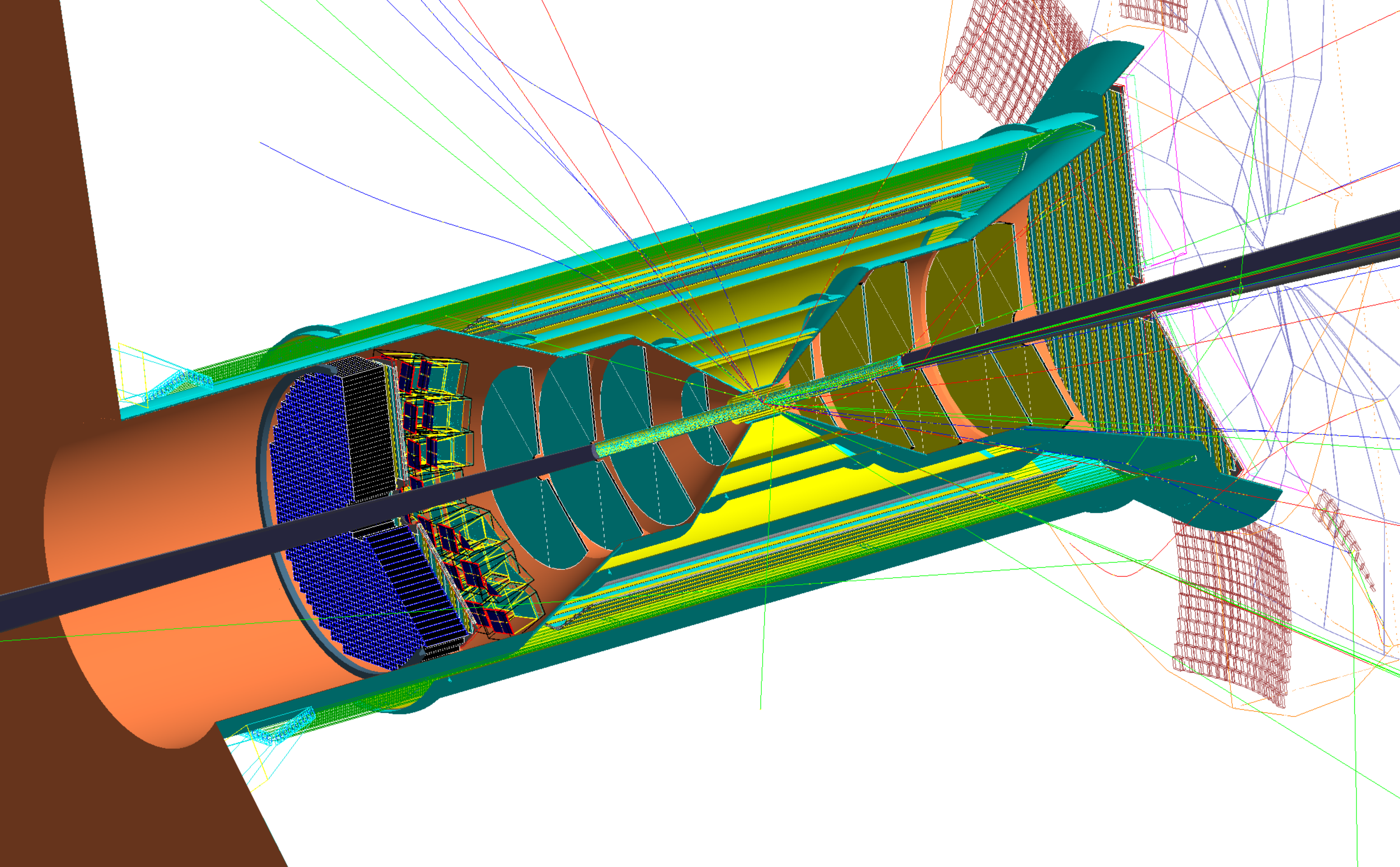}}
\caption{Schematics of the ECCE tracking detector conceptual design. The silicon vertex/tracking sub-detector consists of 5 barrel layers, 5 disks in the hadron beam going direction and 4 disks in the electron beam going direction. Detailed geometry parameters are listed in Table~\ref{tab:si_barrel_geo} and Table~\ref{tab:si_had_geo}.}
\label{fig:ecce_trk_geo}
\end{figure}

A silicon vertex/tracking subsystem \cite{ecce-paper-det-2022-03} with the pseudorapidity coverage of $-3.5 < \eta < 3.5$ has been implemented in the ECCE detector conceptual design (see Figure~\ref{fig:ecce_trk_geo}). The ECCE silicon vertex/tracking detector is based on the Monolithic Active Pixel Sensor (MAPS) technology \cite{maps1}, which is under new developments towards lower material budgets and finer pixel sizes. This technology will determine the primary and displaced vertex for heavy flavor particles. The pixel size is set at 10 micron ($\mu$m) in these simulation studies according to the EIC yellow report. The ECCE silicon barrel sub-detector inherited from the EIC silicon consortium design \cite{Arrington:2021yeb} and the geometry parameters are shown in Table~\ref{tab:si_barrel_geo}. The ECCE hadron and electron endcap silicon tracking sub-detector incorporates the proposed Forward Silicon Tracker (FST) detector design by the Los Alamos National Laboratory EIC team \cite{Li:2020sru, Wong:2020xtc, Li:2021xig, Li:2021kus}. Detailed geometry parameters of the ECCE silicon hadron and electron endcap detector are shown in Table~\ref{tab:si_had_geo}. This silicon vertex/tracking configuration is the default setup for the following heavy flavor hadron and jet reconstruction in simulations. 

Besides the silicon vertex/tracking detector, Micro-Pattern Gas Detector (MPGD) \cite{mpdg_rd51} based gas tracker and AC Coupled Low Gain Avalanche Detector (AC-LGAD) \cite{Giacomini:2019kqz, Heller:2022aug} based outer tracker also contribute to precise tracking momentum determination. The ECCE detector conceptual design has been through several iterations of optimizations and updates, which include integration with different ECCE subsystem designs and layouts, implementation of the associated service parts and mechanical structures \cite{ecce-paper-det-2022-01}. The ECCE tracking detector design has different geometry configurations and technology options \cite{ecce-paper-det-2022-03}, the associated tracking performances are used to evaluate the systematic uncertainties for heavy flavor reconstructions.

\begin{table}[!htb]
\begin{center} 
\scalebox{0.82}{
\begin{tabular}{|c|c|c|c|c|c| } 
 \hline
 Layer index & radius & minimum z & maximum z & pixel pitch \\ \hline
1 & 3.3 cm & -13.5 cm & 13.5 cm & 10 $\mu$m \\ \hline
2 & 4.35 cm & -13.5 cm & 13.5 cm & 10 $\mu$m \\ \hline
3 & 5.4 cm & -13.5 cm & 13.5 cm & 10 $\mu$m \\ \hline
4 & 21.0 cm & -27 cm & 27 cm & 10 $\mu$m \\ \hline
5 & 22.68 cm & -30 cm & 30 cm & 10 $\mu$m \\ \hline
\end{tabular}
 }
 \caption{The ECCE Monolithic Active Pixel Sensor (MAPS) based silicon barrel detector conceptual design geometry.}
    \label{tab:si_barrel_geo}
\end{center}
\end{table}

\begin{table}[!htb]
\begin{center} 
\scalebox{0.87}{
\begin{tabular}{|c|c|c|c|c|c| } 
 \hline
 Disk index & z location & inner radius & outer radius & pixel pitch \\ \hline
hadron 1 & 25 cm & 3.5 cm & 18.5 cm & 10 $\mu$m \\ \hline
hadron 2 & 49 cm & 3.5 cm & 36.5 cm & 10 $\mu$m \\ \hline
hadron 3 & 73 cm & 4.5 cm & 40.5 cm & 10 $\mu$m \\ \hline
hadron 4 & 106 cm & 5.5 cm & 41.5 cm & 10 $\mu$m \\ \hline
hadron 5 & 125 cm & 7.5 cm & 43.5 cm & 10 $\mu$m \\ \hline
electron 1 & -25 cm & 3.5 cm & 18.5 cm & 10 $\mu$m \\ \hline
electron 2 & -52 cm & 3.5 cm & 36.5 cm & 10 $\mu$m \\ \hline
electron 3 & -79 cm & 4.5 cm & 40.5 cm & 10 $\mu$m \\ \hline
electron 4 & -106 cm & 5.5 cm & 41.5 cm & 10 $\mu$m \\ \hline
\end{tabular}
 }
 \caption{The ECCE Monolithic Active Pixel Sensor (MAPS) based silicon hadron and electron endcap detector conceptual design geometry.}
    \label{tab:si_had_geo}
\end{center}
\end{table}


\begin{figure*}[!ht]
\centering
\includegraphics[width=0.86\textwidth]{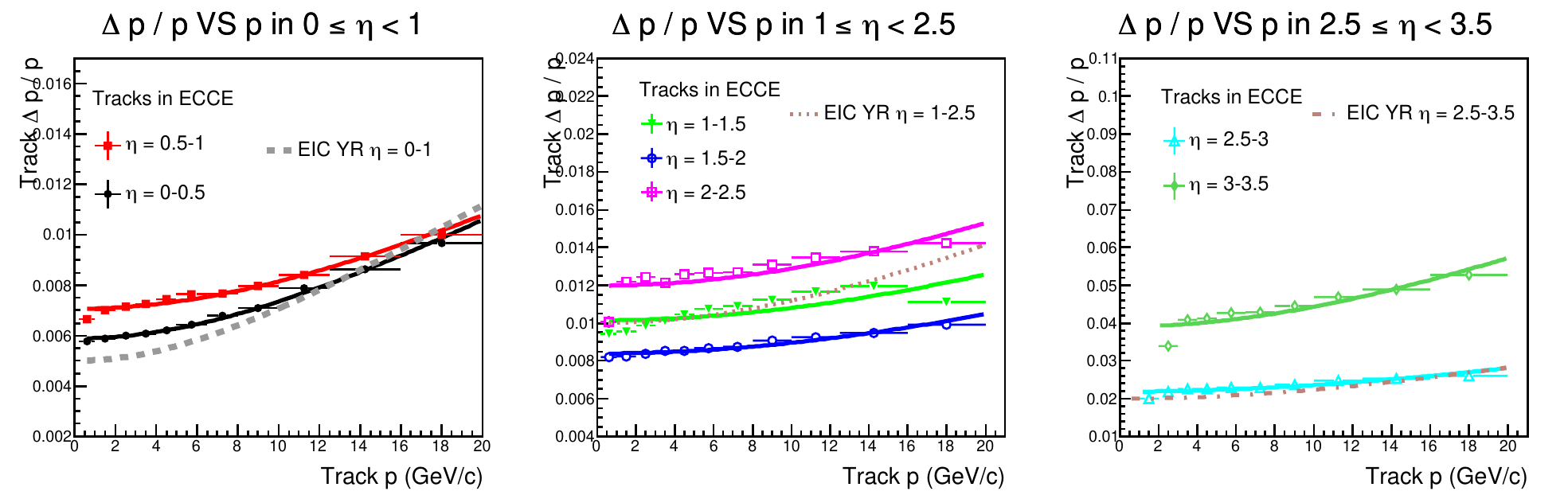}
\includegraphics[width=0.86\textwidth]{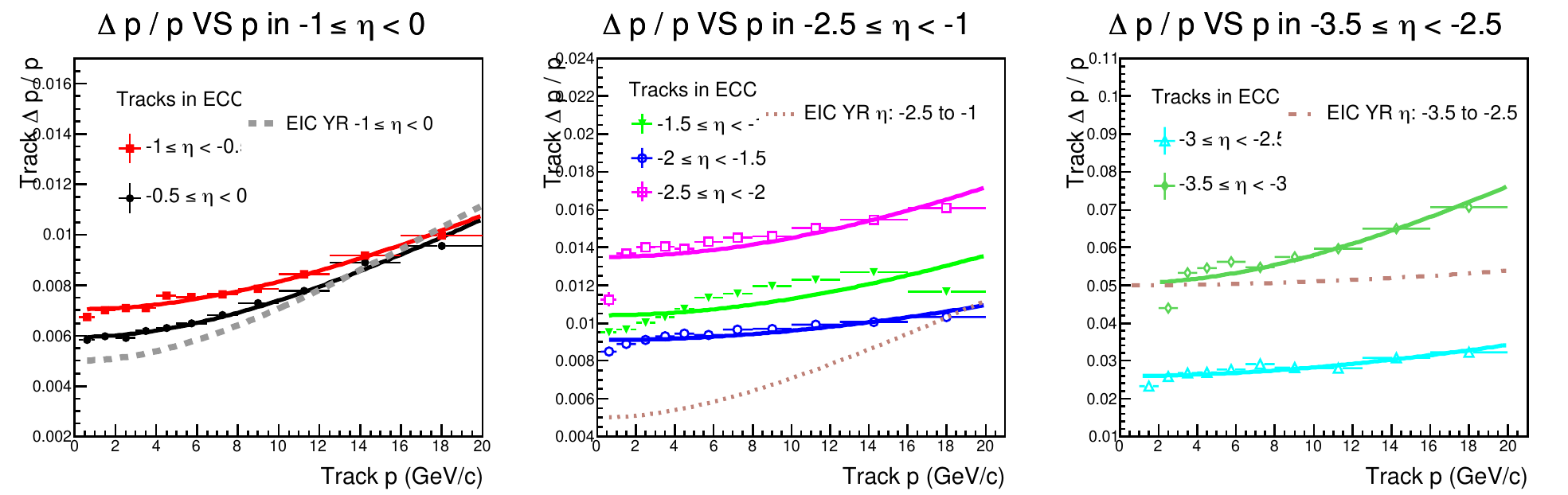}
\caption{Track momentum dependent momentum resolutions in $0 \leq \eta \leq 3.5$ and $-3.5 \leq \eta < 0$ pseudorapidity regions with the ECCE conceptual detector design and the 1.4~T Babar magnet.}
\label{fig:ecce_trk1}
\end{figure*}

\begin{figure*}[!ht]
\centering
\includegraphics[width=0.86\textwidth]{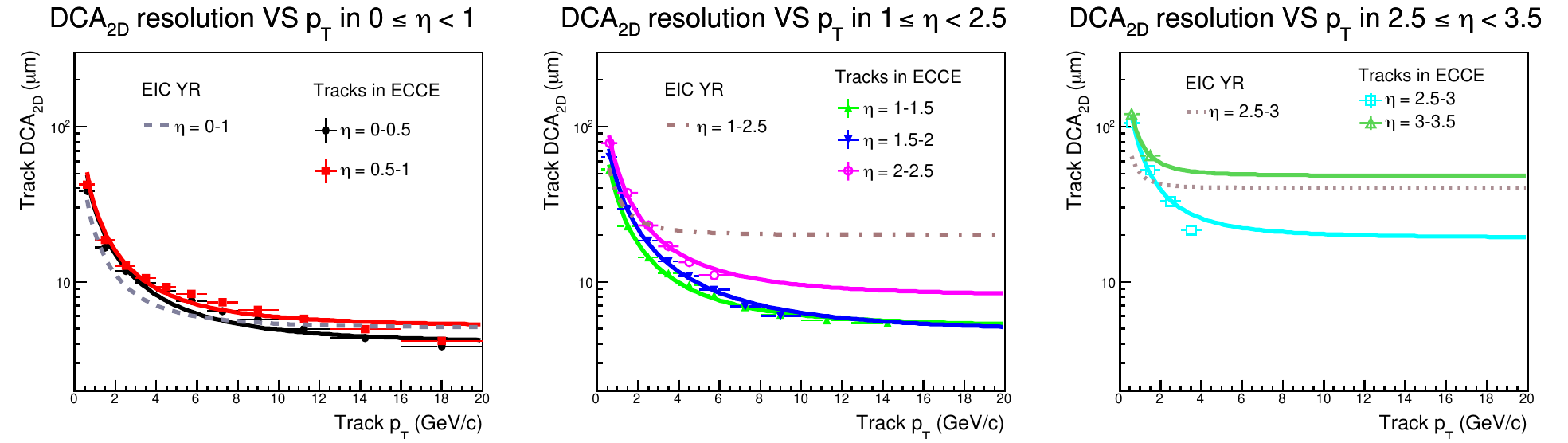}
\includegraphics[width=0.86\textwidth]{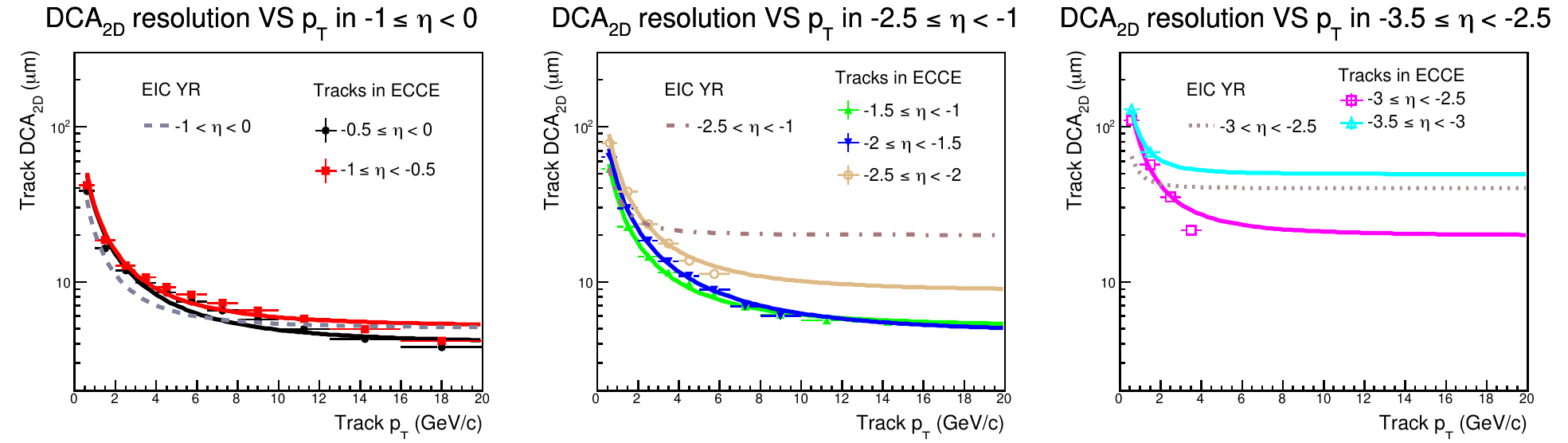}
\caption{Track transverse momentum dependent transverse Distance of Closet Approach ($DCA_{2D}$) resolutions in $0 \leq \eta \leq 3.5$ and $-3.5 \leq \eta < 0$ pseudorapidity regions with the ECCE conceptual detector design and the 1.4~T Babar magnet.}
\label{fig:ecce_trk2}
\end{figure*}

Through single charged pion simulations which are separated in fine pseudorapidity bins, the tracking momentum dependent momentum resolutions of the ECCE default conceptual design \cite{ecce-paper-det-2022-01} are shown in Figure~\ref{fig:ecce_trk1} and transverse momentum ($p_{T}$) dependent resolutions of transverse Distance of Closest Approach ($DCA_{2D}$), which is proportional to particle decay length, are shown in Figure~\ref{fig:ecce_trk2}. All the performance distributions are compared to the EIC yellow report requirement which uses a wider bin separation than this selection. The ECCE momentum resolution is comparable to the performance targets in the Yellow Report, except at the largest forward/backward rapidities. In general, ECCE compensates for this by combining information from multiple detectors. For example, in the backwards region the measure of the kinematics of the scattered electron is improved by incorporating the information from the backwards EMCal.

These tracking performances have been characterized by the recommended fit functions from the EIC yellow report \cite{AbdulKhalek:2021gbh}. For the track momentum dependent momentum resolutions, the fit function is defined below:

\begin{equation}
\label{eq1}
  \sigma(\frac{\Delta p}{p}) = A + B \times p,
\end{equation}
where A and B are free parameters determined by the fit.

The fit function to extract the track transverse momentum dependent $DCA_{2D}$ resolutions is defined as below:

\begin{equation}
\label{eq2}
  \sigma(DCA_{2D}) = C \otimes \frac{D}{p_{T}},
\end{equation}
where C and D are free parameters determined by the fit.

After the fitting, the parameterized momentum and $DCA_{2D}$ resolution  functions per pseudorapidity bin have been implemented in the PYTHIA 8 simulation. All charged particles' momentum and Distance of Closest Approach relative to the primary vertex are smeared for heavy flavor reconstructions.

\subsection{Open Heavy Flavor Hadron and Jet Reconstruction}
\label{sec:rec}
To reconstruct heavy flavor hadrons, a list of track cuts have been applied on the smeared outputs from simulation, which is summarized below:

  \begin{itemize}
  \item $DCA_{2D}$ matching between charged tracks within two-sigma separation of the $DCA_{2D}$ resolutions (in general, the $DCA_{2D}$ resolution ($\Delta DCA_{2D}$) is less than $100~\mu m$).
  \item Each charged track transverse momentum ($p_{T}$) is larger than $0.2~GeV/c$.
  \item Reconstructed heavy flavor hadron transverse momentum ($p_{T}$) is larger than $0.5~GeV/c$.
  \item Reconstructed heavy flavor hadron DCA vector and momenta crossing angle $\theta^{*}$, satisfies $cos(\theta^{*})>0.25$.
  \item charged pion, kaon, proton identification efficiency on average at 95$\%$
  \item charged track reconstruction efficiency on average at $95\%$
  \end{itemize}

\begin{figure*}[ht]
\centering
\includegraphics[width=0.9\textwidth]{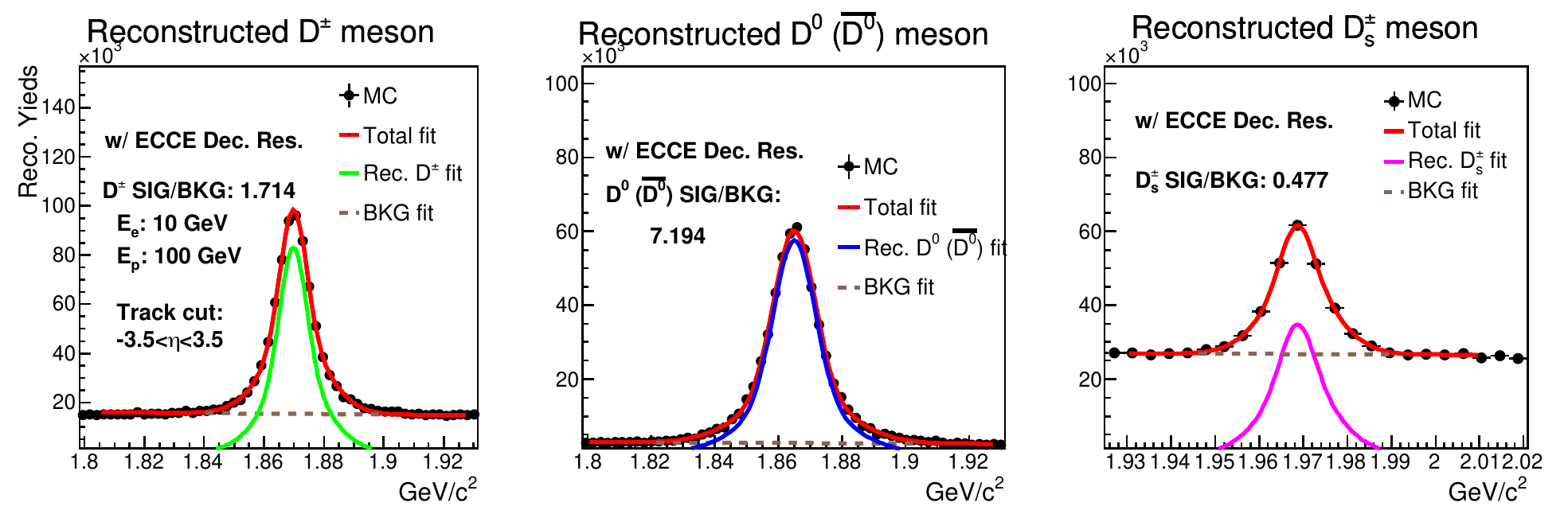}
\includegraphics[width=0.6\textwidth]{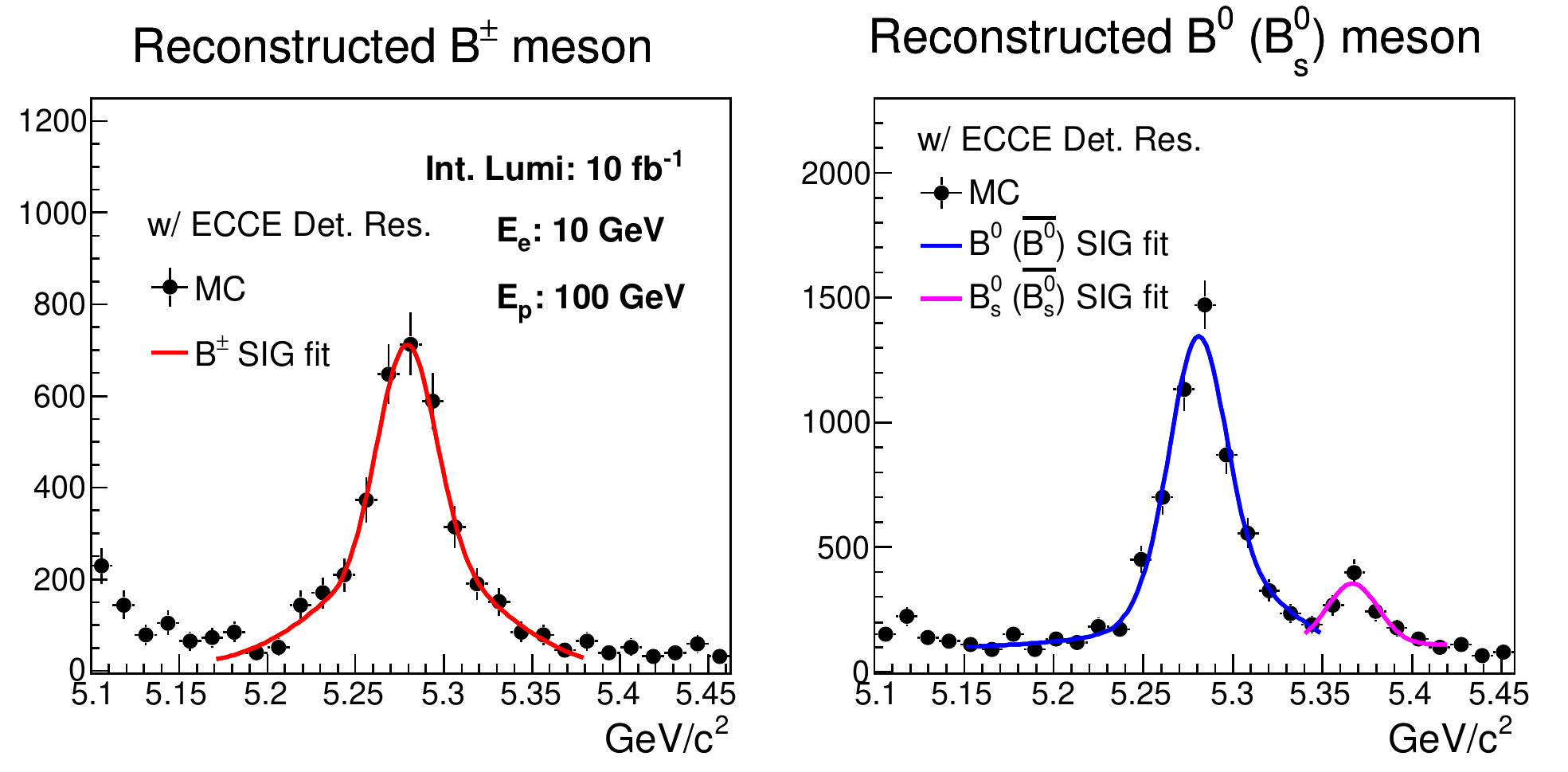}
\caption{Reconstructed heavy flavor hadron mass spectrums with the ECCE reference detector design performance inside the 1.4~T Babar magnet in 10+100 GeV $e+p$ collisions. The top row shows the mass spectrum of reconstructed D mesons and the bottom row show the mass spectrum of reconstructed B mesons. The integrated luminosity is 10 $fb^{-1}$.}
\label{fig:HF_had}
\end{figure*}

\begin{figure*}[!ht]
\centering
\includegraphics[width=0.9\textwidth]{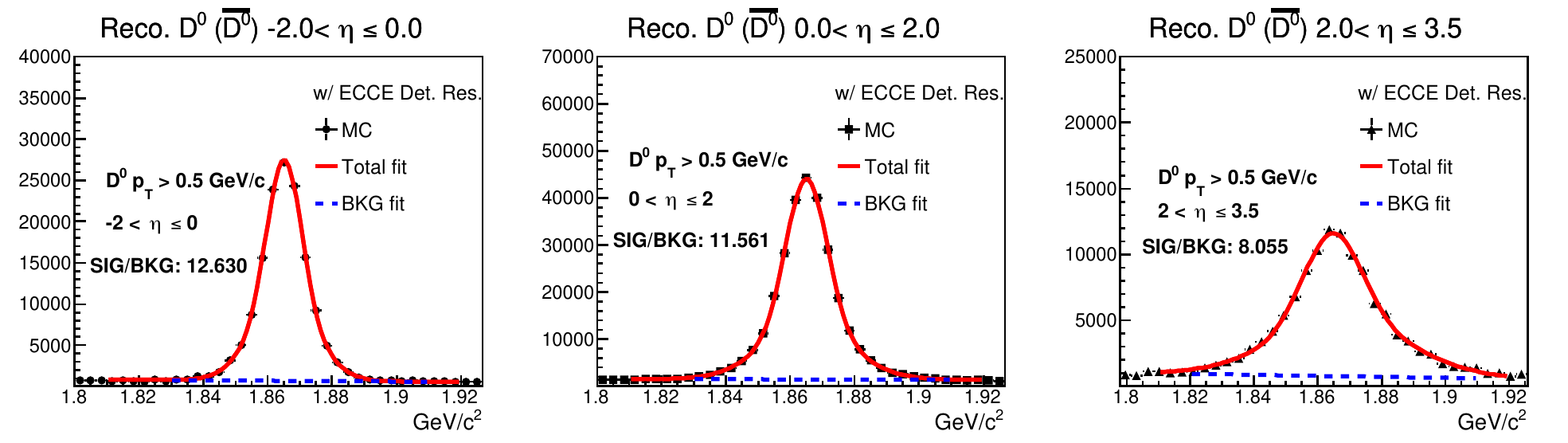}
\caption{Reconstructed $D^{0}$ ($\bar{D^{0}}$) mass distributions in $-2 < \eta \leq 0$ (left), $0 < \eta \leq 2$ (middle), and $2 < \eta \leq 3.5$ (right) pseudorapidity regions with the ECCE detector performance and the 1.4~T Babar magnet in 10+100 GeV $e+p$ collisions. The integrated luminosity is 10 $fb^{-1}$.}
\label{fig:rec_D0}
\end{figure*}

After several iterations of fine tuning of the topological cuts for different heavy flavor hadrons and according to their associated decay channels, a series of reconstructed heavy flavor hadrons have been successfully reconstructed in 10+100 GeV $e+p$ simulation. As illustrated in Figure~\ref{fig:HF_had}, clear signals of reconstructed $D^{\pm}$, $D^{0} (\bar{D^{0}})$, $D_{s}^{\pm}$, $D_{s}^{\pm}$, $B^{\pm}$, $B^{0} (\bar{B^{0}})$ and $B_{s}^{0} (\bar{B_{s}^{0}})$ have been found on top of the combinatorial backgrounds. It is challenging to search for $B_{s}^{0} (\bar{B_{s}^{0}})$ signals with the tracking momentum performance using the 1.4~T Babar magnet and only around one year EIC data collection. These optimized cuts have successfully suppressed backgrounds especially for $D^{0} (\bar{D^{0}})$, which is dominated by two particle decay. The tremendous statistics of inclusive reconstructed D meson allow us to further divide the sample of reconstructed $D^{0}$ ($\bar{D^{0}}$) into four different pseudorapidity regions. Figure~\ref{fig:rec_D0} shows the mass distributions of reconstructed $D^{0}$ ($\bar{D^{0}}$) within $-2 < \eta \leq 0$ (left), $0 < \eta \leq 2$ (middle), and $2 < \eta \leq 3.5$ (right) pseudorapidity regions in 10+100 GeV e+p simulation and projected with $10~fb^{-1}$ integrated luminosity. Good signal over background ratios (as shown in Figure~\ref{fig:rec_D0}) have been achieved in all pseudorapidity regions.

\begin{figure*}[ht]
\centering
\includegraphics[width=0.8\textwidth]{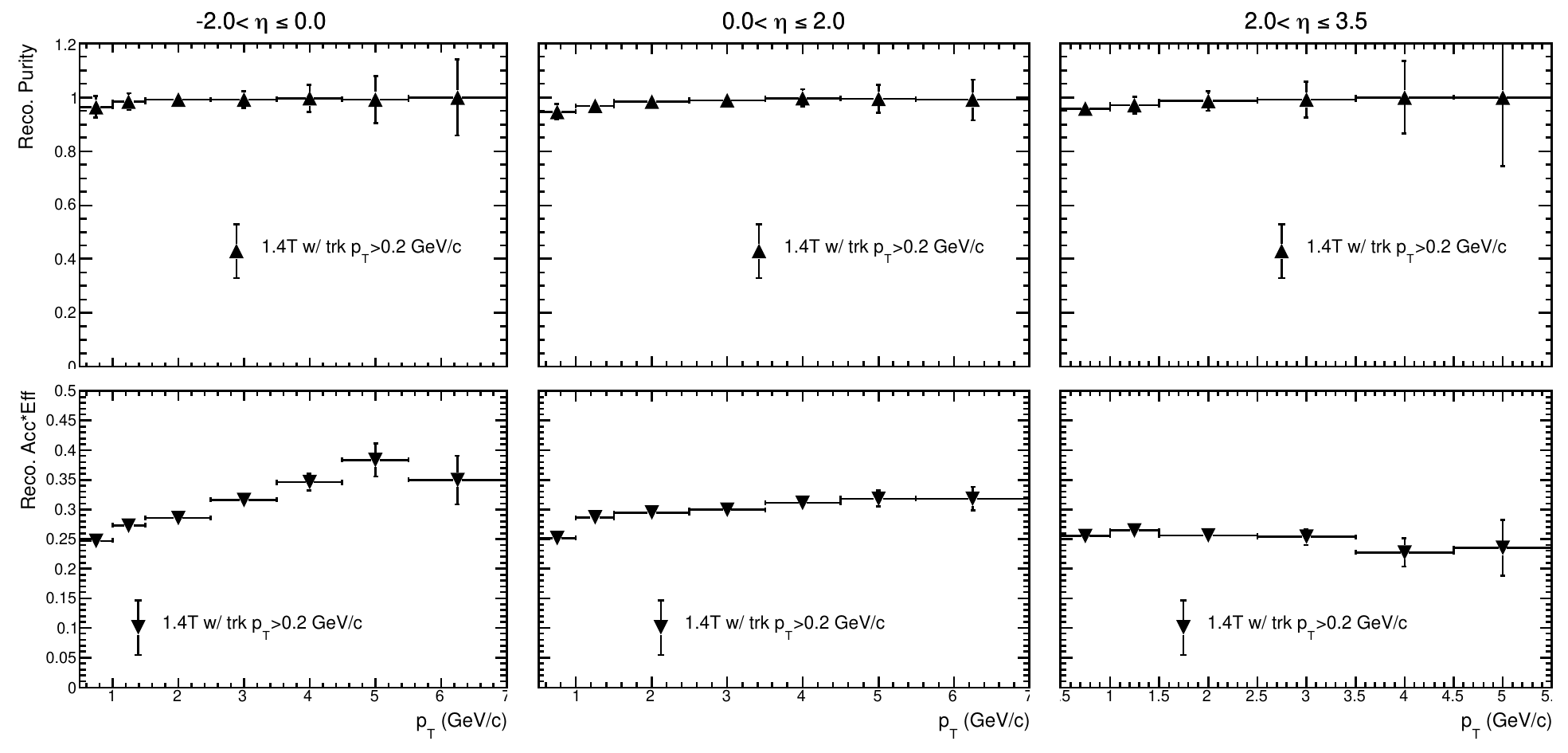}
\caption{Transverse momentum $p_{T}$ dependent purity (top) and acceptance$\times$efficiency (bottom) for reconstructed $D^{0}$ ($\bar{D^{0}}$) in three different pseudorapidity regions with the ECCE detector performance in 10+100 GeV $e+p$ collisions. The transverse momentum for reconstructed $D^{0}$ ($\bar{D^{0}}$) is within the 0.5 GeV/c to 7 GeV/c region. The integrated luminosity is 10 $fb^{-1}$.}
\label{fig:rec_purity_eff1}
\end{figure*}

\begin{figure*}[!ht]
\centering
\includegraphics[width=0.8\textwidth]{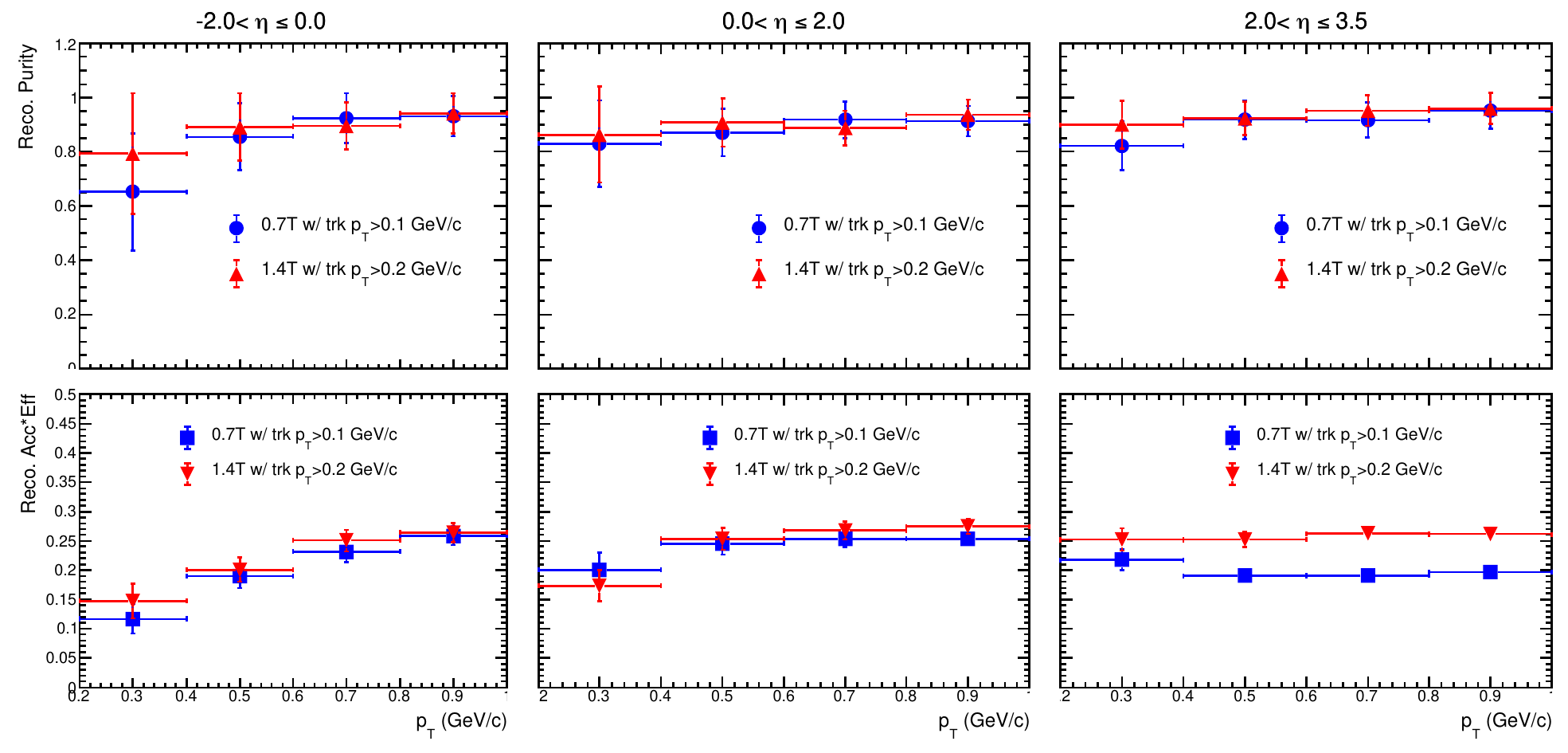}
\caption{Transverse momentum $p_{T}$ dependent purity (top) and acceptance$\times$efficiency (bottom) for reconstructed $D^{0}$ ($\bar{D^{0}}$) in three different pseudorapidity regions with the ECCE detector performance in 10+100 GeV $e+p$ collisions. The transverse momentum for reconstructed $D^{0}$ ($\bar{D^{0}}$) is within 0.2 GeV/c to 1 GeV/c region. The integrated luminosity is 10 $fb^{-1}$.}
\label{fig:rec_purity_eff2}
\end{figure*}

The $D^{0}$ ($\bar{D^{0}}$) reconstruction purity and the acceptance$\times$efficiency have been evaluated in three different pseudorapidity regions, which are $-2 < \eta \leq 0$, $0 < \eta \leq 2$ and $2 < \eta \leq 3.5$. The reconstruciton purity is defined as the yields of reconstructed $D^{0}$ ($\bar{D^{0}}$) associated with true particles divided by the reconstructed $D^{0}$ ($\bar{D^{0}}$) yields. The acceptance$\times$efficiency is defined as the reconstructed $D^{0}$ ($\bar{D^{0}}$) yields divided by the generated $D^{0}$ ($\bar{D^{0}}$) yields in the same kinematic regions. The reconstruction purity and acceptance$\times$efficiency for transverse momentum dependent reconstructed $D^{0}$ ($\bar{D^{0}}$) in e+p simulation are illustrated in Figure~\ref{fig:rec_purity_eff1}. The distributions are studied with track transverse momentum within the $0.5~GeV/c < p_{T} < 7~GeV/c$ window and separated into three different pseudorapidity regions: $-2 < \eta \leq 0$, $0 < \eta \leq 2$, $2 < \eta \leq 3.5$. Figure~\ref{fig:rec_purity_eff2} presents the transverse momentum dependent purity and acceptance$\times$efficiency for reconstructed $D^{0}$ ($\bar{D^{0}}$) with $0.2~GeV/c < p_{T} < 1.0~GeV/c$ in the related pseudorapidity regions as shown in Figure~\ref{fig:rec_purity_eff1}. Values of these reconstruction acceptance$\times$efficiency are mainly due to the $D^{0}$ ($\bar{D^{0}}$) decay kinematics and cut selections. 

\begin{figure}[!ht]
\centering
\includegraphics[width=0.4\textwidth]{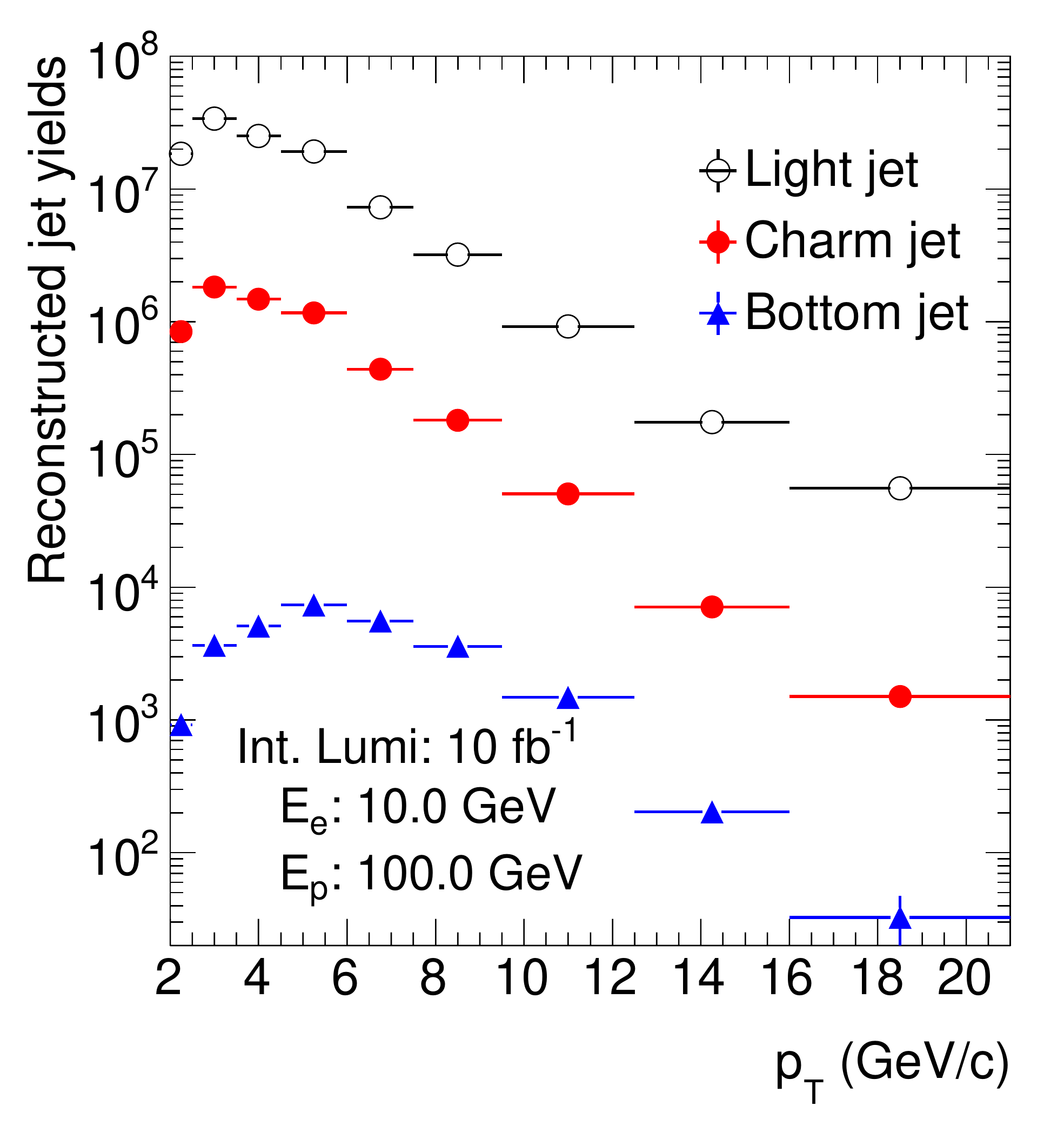}
\caption{Reconstructed heavy flavor jet transverse momentum $p_{T}$ distributions with the ECCE detector performance and 1.4 T Babar magnet in 10+100 GeV $e+p$ collisions. The reconstructed light jet yields are shown in black open circles, distributions for charm jets and bottom jets shown in red closed circles and blue closed triangles, respectively. The integrated luminosity is 10 $fb^{-1}$.}
\label{fig:HF_jet}
\end{figure}

In addition to heavy flavor hadron reconstruction, we have carried out the charm and bottom jet studies in $e+p$ simulations. For jet reconstruction, particle energy smearing based on the proposed ECCE ElectroMagnetic Calorimeter (EMCal) and Hadronic Calorimeter (HCal) responses \cite{ecce-paper-det-2022-01} have been included in simulations. The anti-$k_{T}$ jet cone radius R ($\sqrt{(\eta_{jet}-\eta_{track})^2+(\phi_{jet}-\phi_{track })^2}$) is selected as 1.0 as the particle multiplicity is lower in $e+p/A$ collisions compared to $p+p/A+A$ collisions. 

For heavy flavor jet tagging, we started with heavy flavor hadron and jet association requirement, which is at least a fully reconstructed heavy flavor hadron is within the jet cone, to tag heavy flavor jets. Initial work has been preformed in \cite{Li:2020wyc}, which depends on whether a jet contains a fully reconstructed charm or bottom hadron. Due to the low tagging efficiency, we switched to the displaced/decay vertex tagging method \cite{CMS:2017wtu,ATLAS:2015thz}, which searches for displaced/decay vertex reconstructed from tracks within a jet and matches to a charm or bottom hadron decay. The reconstructed displaced vertex is required to be associated with at least two tracks inside the jet. Through this new tagging method, the tagging efficiency improves significantly, which has been reflected in the reconstructed jet yields. If a jet does not contain a displaced vertex associated with a heavy flavor product decay, then this jet is marked as a light flavor jet.

Figure~\ref{fig:HF_jet} illustrates the jet transverse momentum dependent yields of reconstructed jets with different flavors in 10+100~$e+p$ collisions with integrated luminosity at 10~$fb^{-1}$. The jet reconstruction efficiencies have been included in these yields. The jet purity is greater than 60$\%$ and varies with jet transverse momentum. We plan to apply novel machine learning algorithms to further enhance the jet reconstruction purity and efficiency.

\section {Results}
\label{sec:results}
The projected statistical accuracy of reconstructed heavy flavor hadron and jet cross sections have been studied in $e+p$ simulations. As an $e+A$ event generator is still in development, the heavy flavor hadron/jet spectrums in $e+Au$ collisions are obtained through the nucleus mass number (A) scaled $e+p$ cross section at the same collision energy.

\subsection{Flavor dependent inclusive heavy flavor hadron nuclear modification factor $R_{eAu}$ projection}
\label{sec:result1}
\begin{figure}[!ht]
\centering
\includegraphics[width=0.42\textwidth]{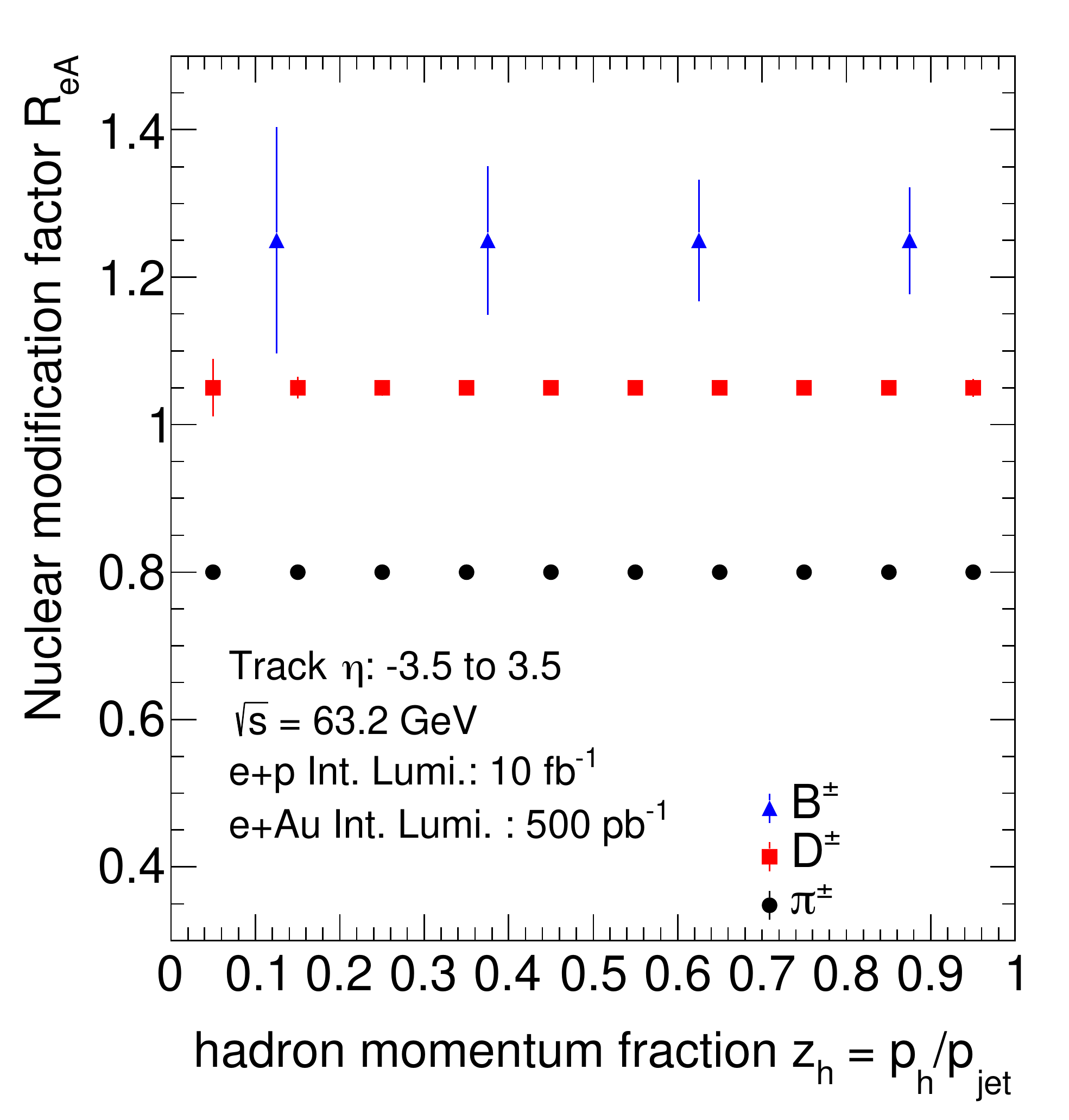}
\caption{Projected statistical uncertainties of reconstructed hadron momentum fraction $z_{h}$ dependent nuclear modification factor $R_{eAu}$ for $\pi^{\pm}$, $D^{\pm}$ and $B^{\pm}$ with the ECCE detector performance in 10+100 GeV $e+p$ and $e+Au$ collisions. The tracking performance is evaluated with the 1.4T Babar magnet. The integrated luminosity for $e+p$ ($e+Au$) collisions is 10 $fb^{-1}$ (500 $pb^{-1}$).}
\label{fig:ReA_babar}
\end{figure}

One key kinematic parameter of the fragmentation function is the hadron momentum fraction $z_{h}$, which is defined as the final state hadron momentum fraction relative to the initial quark/gluon. In experiments, the hadron momentum fraction is measured as the ratio of reconstructed hadron momentum over the associated jet momentum. The flavor dependent inclusive heavy flavor hadron nuclear modification factor $R_{eAu}$ versus hadron momentum fraction have been studied with reconstructed heavy flavor hadron and jet yields using the ECCE detector performances, which have been specifically discussed in section~\ref{sec:rec}. Correlations between reconstructed charm (bottom) hadrons and charm (bottom) jets have been carried out through requiring the reconstructed charm (bottom) hadrons to be within the charm (bottom) jet cone ($\Delta R_{h-jet} \equiv \sqrt{(\eta_{h} - \eta_{jet})^{2}+(\varphi_{h}- \varphi_{jet})^{2}} < 1.0$). After this selection, the hadron momentum fraction for charm hadrons is defined as the reconstructed charm hadron momentum over the reconstructed charm jet momentum. The same definition has been applied for reconstructed bottom hadrons and jets respectively. The projected statistical uncertainties of hadron momentum fraction $z_{h}$ dependent nuclear modification factor $R_{eAu}$ with the ECCE detector performance for reconstructed $\pi^{\pm}$, $D^{\pm}$ mesons and $B^{\pm}$ mesons in 10+100 GeV $e+p$ and $e+Au$ collisions are shown in Figure~\ref{fig:ReA_babar}. The integrated luminosity for $e+p$ collisions is 10 $fb^{-1}$ and the value for $e+Au$ is 500 $pb^{-1}$.

\begin{figure*}[!ht]
\centering
\includegraphics[width=0.9\textwidth]{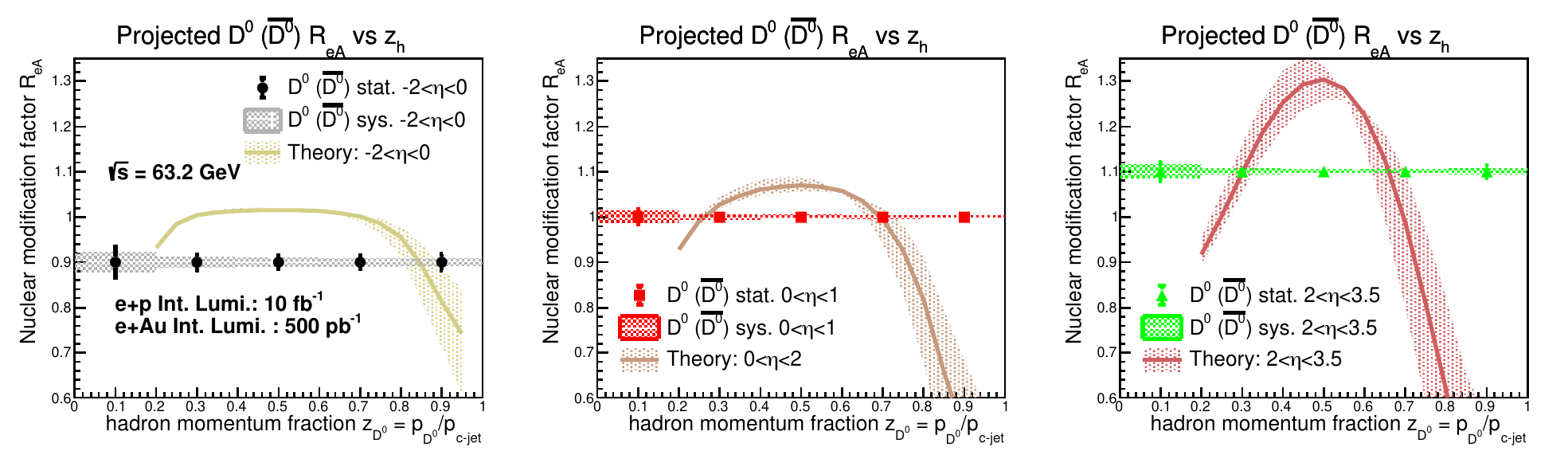}
\caption{Hadron momentum fraction $z_{h}$ dependent projected uncertainties of nuclear modification factor $R_{eAu}$ for reconstructed $D^{0}$ ($\bar{D^{0}}$) with the ECCE detector performance in 10+100 GeV $e+p$ and $e+Au$ collisions. The integrated luminosity for $e+p$ ($e+Au$) collisions is 10 $fb^{-1}$ (500 $pb^{-1}$). The systematical uncertainties come from different detector designs, magnet options and jet cone selections. The theoretical calculations are from \cite{Li:2020zbk}.}
\label{fig:eta_ReA_Dmeson}
\end{figure*}

\begin{figure*}[!ht]
\centering
\includegraphics[width=0.9\textwidth]{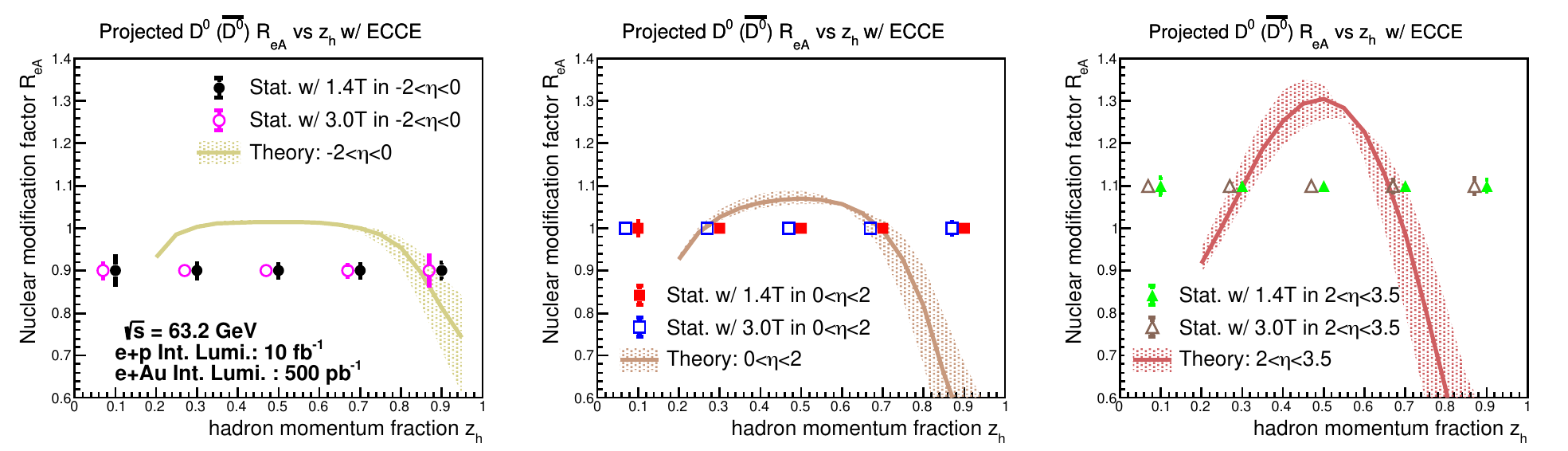}
\caption{Hadron momentum fraction $z_{h}$ dependent projected uncertainties of nuclear modification factor $R_{eAu}$ for reconstructed $D^{0}$ ($\bar{D^{0}}$) with the ECCE detector performance in 10+100 GeV $e+p$ and $e+Au$ collisions. The integrated luminosity for $e+p$ ($e+Au$) collisions is 10 $fb^{-1}$ (500 $pb^{-1}$). The statistical uncertainties of the projected $R_{eAu}$ with the ECCE detector performance using the 1.4 T Babar (3.0 T Beast) magnet are shown in closed (open) markers. The theoretical calculations are from \cite{Li:2020zbk}.}
\label{fig:eta_ReA_Dmeson2}
\end{figure*}


\subsection{Pseudorapidity dependent reconstructed D meson nuclear modification factor $R_{eAu}$ projection}
\label{sec:result2}

Heavy flavor productions in different pseudorapidity regions have different sensitivities to access the initial-state and final-state effects in vacuum and nuclear medium. Recent theoretical developments based on the energy loss mechanism have extensively calculated the nuclear modification factor for D-meson and B-meson in $e+Au$ collisions at different center of mass energies and different pseudorapidities \cite{Li:2020zbk}. Abundant statistics can be achieved for charm hadron/jet reconstruction at the EIC as demonstrated in section~\ref{sec:rec}. With performances of the ECCE conceptual detector design, the projected statistical and systematical uncertainties of hadron momentum fraction $z_{h}$ dependent nuclear modification factor $R_{eAu}$ for reconstructed $D^{0}$ ($\bar{D^{0}}$) within the pseudorapidity bins of $-2 < \eta \leq 0$, $0 < \eta \leq 2$ and $2 < \eta \leq 3.5$ in 10+100 GeV $e+p$ and $e+Au$ collisions are illustrated in Figure~\ref{fig:eta_ReA_Dmeson}. The systematical uncertainties in these proposed measurements include reconstructed $D^{0}$ ($\bar{D^{0}}$) yield variations from different ECCE detector designs \cite{ecce-paper-det-2022-01} and jet cone radius selections (R=0.7 to 1.0). These studies are compared to theoretical calculations \cite{Li:2020zbk}, which are based on the parton energy loss model and scaled by inclusive hadron yields in the respective pseudorapidity bins. These theoretical and experimental comparisons indicate that the future EIC heavy flavor hadron and jet measurements can provide great precision in extracting the fragmentation function in different nuclear medium sizes within the access kinematic coverage especially at high hadron momentum fraction region and forward pseudorapidities, which are little known.

 Figure~\ref{fig:eta_ReA_Dmeson2} compares the projected uncertainties of hadron momentum fraction $z_{h}$ dependent nuclear modification factor $R_{eAu}$ for reconstructed $D^{0}$ ($\bar{D^{0}}$) in 10+100 GeV $e+p$ and $e+Au$ collisions with the ECCE detector performance with Babar and Beast magnet options in the pseudorapidity regions of $-2 < \eta \leq 0$, $0 < \eta \leq 2$ and $2 < \eta \leq 3.5$. As presented in Figure~\ref{fig:eta_ReA_Dmeson} and Figure~\ref{fig:eta_ReA_Dmeson2}, better experimental precision can be obtained with the ECCE detector design for reconstructed $D^{0}$ ($\bar{D^{0}}$) $R_{eAu}$ projections compared to the theoretical uncertainties especially in the forward rapidity regions. 
 
 As the reconstructed hadron pseudorapidity changes from backward going direction to the forward going direction, the projected $D^{0}$ ($\bar{D^{0}}$) $R_{eAu}$ based on the parton energy loss model tends to change from a suppression to a enhancement. Unlike the parton energy loss model, the absorption model indicates that the heavy flavor $R_{eA}$ should be less than 1. With the precision provided by the ECCE detector design, good discriminating power will be provided by the proposed open heavy flavor hadron $R_{eA}$ measurements in separating different model predictions in exploring the hadronization process in nuclear medium.

\section {Summary and Outlook}
\label{sec:summary}

Detailed simulation studies have been carried out for open heavy flavor hadron and jet reconstruction and physics projections with the ECCE detector performance and latest EIC accelerator design. Great signatures have be observed for D-meson, B-meson, charm and bottom jet productions in $e+p$ simulations, which has validated the performance of the ECCE detector, especially the ECCE tracking detector subsystem can deliver the desired EIC physics. The proposed heavy flavor hadron and jet measurements at the EIC will play a critical role in exploring the heavy quark hadronization process in different sizes of nuclear medium with great precision and a broad kinematic coverage. The EIC detector developments will advance from the reference detector design: the ECCE detector conceptual design, new physics observables such as the the heavy flavor hadron and jet correlations and heavy flavor jet substructure will be studied with new EIC detector design and performance.
\section{Acknowledgements}
\label{acknowledgements}

We thank the EIC Silicon Consortium for cost estimate methodologies concerning silicon tracking systems, technical discussions, and comments.  We acknowledge the important prior work of projects eRD16, eRD18, and eRD25 concerning research and development of MAPS silicon tracking technologies.

We thank the EIC LGAD Consortium for technical discussions and acknowledge the prior work of project eRD112.

We thank (list of individuals who are not coauthors) for their useful discussions, advice, and comments.

We acknowledge support from the Office of Nuclear Physics in the Office of Science in the Department of Energy, the National Science Foundation, and the Los Alamos National Laboratory Laboratory Directed Research and Development (LDRD) 20200022DR.


\bibliographystyle{elsarticle-num} 
\bibliography{refs.bib,refs-ecce.bib}

\begin{thebibliography}{10}
\expandafter\ifx\csname url\endcsname\relax
  \def\url#1{\texttt{#1}}\fi
\expandafter\ifx\csname urlprefix\endcsname\relax\def\urlprefix{URL }\fi
\expandafter\ifx\csname href\endcsname\relax
  \def\href#1#2{#2} \def\path#1{#1}\fi

\bibitem{Accardi:2012qut}
A.~Accardi, et~al., {Electron Ion Collider: The Next QCD Frontier}:
  {Understanding the glue that binds us all}, Eur. Phys. J. A 52~(9) (2016)
  268.
\newblock \href {http://arxiv.org/abs/1212.1701} {\path{arXiv:1212.1701}},
  \href {https://doi.org/10.1140/epja/i2016-16268-9}
  {\path{doi:10.1140/epja/i2016-16268-9}}.

\bibitem{NAP25171}
{National Academies of Sciences, Engineering, and Medicine},
  \href{https://www.nap.edu/catalog/25171/an-assessment-of-us-based-electron-ion-collider-science}{An
  Assessment of U.S.-Based Electron-Ion Collider Science}, The National
  Academies Press, Washington, DC, 2018.
\newblock \href {https://doi.org/10.17226/25171} {\path{doi:10.17226/25171}}.
\newline\urlprefix\url{https://www.nap.edu/catalog/25171/an-assessment-of-us-based-electron-ion-collider-science}

\bibitem{AbdulKhalek:2021gbh}
R.~Abdul~Khalek, et~al., {Science Requirements and Detector Concepts for the
  Electron-Ion Collider: EIC Yellow Report} (3 2021).
\newblock \href {http://arxiv.org/abs/2103.05419} {\path{arXiv:2103.05419}}.

\bibitem{Collins:1985ue}
J.~C. Collins, D.~E. Soper, G.~F. Sterman, {Factorization for Short Distance
  Hadron - Hadron Scattering}, Nucl. Phys. B 261 (1985) 104--142.
\newblock \href {https://doi.org/10.1016/0550-3213(85)90565-6}
  {\path{doi:10.1016/0550-3213(85)90565-6}}.

\bibitem{ecce-paper-det-2022-03}
{First Authors, et al.}, {Design and Simulated Performance of Tracking Systems
  for the ECCE Detector at the Electron Ion Collider}, to be published in Nucl.
  Instrum. Methods A (in this issue) (2022).

\bibitem{Sjostrand:2014zea}
T.~Sj\"ostrand, S.~Ask, J.~R. Christiansen, R.~Corke, N.~Desai, P.~Ilten,
  S.~Mrenna, S.~Prestel, C.~O. Rasmussen, P.~Z. Skands, {An introduction to
  PYTHIA 8.2}, Comput. Phys. Commun. 191 (2015) 159--177.
\newblock \href {http://arxiv.org/abs/1410.3012} {\path{arXiv:1410.3012}},
  \href {https://doi.org/10.1016/j.cpc.2015.01.024}
  {\path{doi:10.1016/j.cpc.2015.01.024}}.

\bibitem{AGOSTINELLI2003250}
S.~Agostinelli, et~al., Geant4—a simulation toolkit, Nuclear Instruments and
  Methods in Physics Research Section A: Accelerators, Spectrometers, Detectors
  and Associated Equipment 506~(3) (2003) 250--303.
\newblock \href {https://doi.org/https://doi.org/10.1016/S0168-9002(03)01368-8}
  {\path{doi:https://doi.org/10.1016/S0168-9002(03)01368-8}}.

\bibitem{fun4allGithub}
Eic,
  \href{https://github.com/eic/fun4all\_coresoftware}{fun4all\_coresoftware}.
\newline\urlprefix\url{https://github.com/eic/fun4all\_coresoftware}

\bibitem{ecce-note-comp-2021-01}
{ECCE Consortium}, \href{https://www.ecce-eic.org/ecce-internal-notes}{{ECCE
  Computing Plan}}, ecce-note-comp-2021-01 (2021).
\newline\urlprefix\url{https://www.ecce-eic.org/ecce-internal-notes}

\bibitem{maps1}
G.~Deptuch, J.~Berst, G.~Claus, C.~Colledani, W.~Dulinski, U.~Goerlach,
  Y.~Gomoushkin, Y.~Hu, D.~Husson, G.~Orazi, R.~Turchetta, J.~Riester,
  M.~Winter, Design and testing of monolithic active pixel sensors for charged
  particle tracking, in: 2000 IEEE Nuclear Science Symposium. Conference Record
  (Cat. No.00CH37149), Vol.~1, 2000, pp. 3/41--3/48 vol.1.
\newblock \href {https://doi.org/10.1109/NSSMIC.2000.949010}
  {\path{doi:10.1109/NSSMIC.2000.949010}}.

\bibitem{Arrington:2021yeb}
J.~Arrington, et~al., {EIC Physics from An All-Silicon Tracking Detector} (2
  2021).
\newblock \href {http://arxiv.org/abs/2102.08337} {\path{arXiv:2102.08337}}.

\bibitem{Li:2020sru}
X.~Li, et~al., {A New Heavy Flavor Program for the Future Electron-Ion
  Collider}, EPJ Web Conf. 235 (2020) 04002.
\newblock \href {http://arxiv.org/abs/2002.05880} {\path{arXiv:2002.05880}},
  \href {https://doi.org/10.1051/epjconf/202023504002}
  {\path{doi:10.1051/epjconf/202023504002}}.

\bibitem{Wong:2020xtc}
C.-P. Wong, X.~Li, M.~Brooks, M.~J. Durham, M.~X. Liu, A.~Morreale,
  C.~da~Silva, W.~E. Sondheim, {A Proposed Forward Silicon Tracker for the
  Future Electron-Ion Collider and Associated Physics Studies} (9 2020).
\newblock \href {http://arxiv.org/abs/2009.02888} {\path{arXiv:2009.02888}}.

\bibitem{Li:2021xig}
X.~Li, {Heavy Flavor and Jet Studies for the Future Electron-Ion Collider to
  Explore the Hadronization Process}, in: {28th International Workshop on Deep
  Inelastic Scattering and Related Subjects}, 2021.
\newblock \href {http://arxiv.org/abs/2107.09035} {\path{arXiv:2107.09035}},
  \href {https://doi.org/10.21468/SciPostPhysProc.8.076}
  {\path{doi:10.21468/SciPostPhysProc.8.076}}.

\bibitem{Li:2021kus}
X.~Li, M.~Brooks, M.~Durham, Y.~C. Morales, A.~Morreale, C.~Prokop, E.~Renner,
  W.~Sondheim, {Forward silicon vertex/tracking detector design and R$\&$D for
  the future Electron-Ion Collider}, PoS PANIC2021 (2022) 084.
\newblock \href {http://arxiv.org/abs/2111.03182} {\path{arXiv:2111.03182}},
  \href {https://doi.org/10.22323/1.380.0084} {\path{doi:10.22323/1.380.0084}}.

\bibitem{mpdg_rd51}
M.~Titov, L.~Ropelewski, {Micro-Pattern Gaseous Detector Technologies and RD51
  Collaboration}, Modern Physics Letters A 28~(13) (2013) 1340022.
\newblock \href {https://doi.org/10.1142/S0217732313400221}
  {\path{doi:10.1142/S0217732313400221}}.

\bibitem{Giacomini:2019kqz}
G.~Giacomini, W.~Chen, G.~D'Amen, A.~Tricoli, {Fabrication and performance of
  AC-coupled LGADs}, JINST 14~(09) (2019) P09004.
\newblock \href {http://arxiv.org/abs/1906.11542} {\path{arXiv:1906.11542}},
  \href {https://doi.org/10.1088/1748-0221/14/09/p09004}
  {\path{doi:10.1088/1748-0221/14/09/p09004}}.

\bibitem{Heller:2022aug}
R.~Heller, et~al., {Characterization of BNL and HPK AC-LGAD sensors with a 120
  GeV proton beam} (1 2022).
\newblock \href {http://arxiv.org/abs/2201.07772} {\path{arXiv:2201.07772}}.

\bibitem{ecce-paper-det-2022-01}
{First Authors, et al.}, {Design of the ECCE Detector for the Electron Ion
  Collider}, to be published in Nucl. Instrum. Methods A (in this issue)
  (2022).

\bibitem{Li:2020wyc}
X.~Li, {Heavy flavor and jet studies for the future Electron-Ion Collider}, PoS
  HardProbes2020 (2021) 175.
\newblock \href {http://arxiv.org/abs/2007.14417} {\path{arXiv:2007.14417}},
  \href {https://doi.org/10.22323/1.387.0175} {\path{doi:10.22323/1.387.0175}}.

\bibitem{CMS:2017wtu}
A.~M. Sirunyan, et~al., {Identification of heavy-flavour jets with the CMS
  detector in pp collisions at 13 TeV}, JINST 13~(05) (2018) P05011.
\newblock \href {http://arxiv.org/abs/1712.07158} {\path{arXiv:1712.07158}},
  \href {https://doi.org/10.1088/1748-0221/13/05/P05011}
  {\path{doi:10.1088/1748-0221/13/05/P05011}}.

\bibitem{ATLAS:2015thz}
G.~Aad, et~al., {Performance of $b$-Jet Identification in the ATLAS
  Experiment}, JINST 11~(04) (2016) P04008.
\newblock \href {http://arxiv.org/abs/1512.01094} {\path{arXiv:1512.01094}},
  \href {https://doi.org/10.1088/1748-0221/11/04/P04008}
  {\path{doi:10.1088/1748-0221/11/04/P04008}}.

\bibitem{Li:2020zbk}
H.~T. Li, Z.~L. Liu, I.~Vitev, {Heavy meson tomography of cold nuclear matter
  at the electron-ion collider}, Phys. Lett. B 816 (2021) 136261.
\newblock \href {http://arxiv.org/abs/2007.10994} {\path{arXiv:2007.10994}},
  \href {https://doi.org/10.1016/j.physletb.2021.136261}
  {\path{doi:10.1016/j.physletb.2021.136261}}.

\end{thebibliography}

\end{document}